\documentclass[fleqn,10pt]{SelfArx} 

\usepackage{lipsum} 

\definecolor{color1}{RGB}{0,0,90} 
\definecolor{color2}{RGB}{0,20,20} 

\usepackage{hyperref} 
\hypersetup{hidelinks,colorlinks,breaklinks=true,urlcolor=color2,citecolor=color1,linkcolor=color1,bookmarksopen=false,pdftitle={Title},pdfauthor={Author}}

\JournalInfo{Geomagnetism and Aeronomy, 2018, Vol. 58, No. 7} 
\Archive{DOI: 10.1134/S0016793218070095} 

\PaperTitle{
Properties of Kepler Stars with the Most Powerful Flares
} 

\Authors{M.\,M.~Katsova\textsuperscript{1}, B.\,A.~Nizamov\textsuperscript{1,2}
} 

\affiliation{\textsuperscript{1}\textit{
Sternberg State Astronomical Institute, Lomonosov Moscow State University, Moscow, Russia
}} 
\affiliation{\textsuperscript{2}\textit{
Faculty of Physics, Lomonosov Moscow State University, Moscow, Russia
}} 
\affiliation{E-mail: maria@sai.msu.ru, nizamov@physics.msu.ru}
\affiliation{Received: February 25, 2018}

\newcommand{\keywordname}{Keywords} 

\Abstract{
We analyze physical characteristics of late-type stars wherein the Kepler mission registered superflares. We use the revised stellar fundamental parameters, i.e. effective temperatures $T_{\textit{eff}}$ and surface gravitational accelerations $\log g$, from the Kepler archive published in 2017, as compared to previous studies by Balona (2015) based on the release of 2011. Among superflare stars there are both single objects and members of eclipsing binaries. We select the late-type stars (with $T_{\textit{eff}}  < 6500$\,K) wherein occured the most powerful flares with the total flare energy $> 10^{35}$\,erg and consider their locations in the $T_{\textit{eff}} - \log g$ diagram. Both components of binaries and single stars appear to reside mostly in between the main sequence and the subgiant branches and therefore have larger radii compared to that of the Sun. Besides, as a rule these single stars are fast rotators and can be considered as young objects that it is difficult to attribute to “solar-type stars”. Extremely high flare energy of these stars requires quite strong magnetic fields that cannot be generated even due to scaling of the solar dynamo. Apparently, for explanation of the strongest non-stationary phenomena on stars considered, it would be worthwhile to attract another regime of the dynamo mechanism that can be realized in these objects.
}

\begin{document}

\flushbottom 

\maketitle 


\thispagestyle{empty} 

\section{Introduction}

Until recently, data on flares in the main sequence G dwarfs (except the Sun) were practically absent. Only in 2012 there appeared the results of the Kepler space telescope observations which monitored more than 160000 stars. In the dedicated G dwarf investigation (Maehara et al., 2012, Shibayama et al., 2013), the discovery of powerful flares with energies in excess of $10^{33}$\,erg is presented; it is based on the observations of almost 83000 stars where 365 superflares were detected in 120 days in 148 solar type stars. The main Kepler mission continued from April, 2009 until May, 2013. Its archive contains the data on continuous photometry during this period with the time resolution of 30 min and 1 min. The observations with the 1-min cadence are suitable for the search of flares. From these data, Maehara et al. (2015) selected 1547 single solar type stars with the effective temperature in the range $5300\,K < T_{\textit{eff}} < 6300$\,K and $\log g$ in the range $4.0 < \log g < 4.8$. Just in 23 solar type stars they found 187 flares with the total energy from 
$2\times 10^{32}$ to $8\times 10^{35}$\,erg. The flare occurrence rate depends on the axial rotation period of these stars. More often flares occur in fast rotating stars 
($P_{rot} \le 3–5$\,days). In stars with the periods between 10 and 20 days flare occurrence rate is approximately 8 times higher than in those with longer periods. In particular, flares with the total energy $E > 10^{33}$\,erg in slowly rotating stars are almost absent. According to Maehara et al. (2015), mean occurrence rate of flares with the energy of 
$10^{33}$\,erg on a solar-type star is one event in approximately 70 years, with the energy of $10^{34}$\,erg -- one in 500 years, with the energy of $10^{35}$ years -- one in 4000 years. These authors estimate the mean rate of X100 flares on a star with the period of 25 days, like in the Sun, as one event in 500 -- 600 years.

It is worth noting that the majority of superflare stars demonstrate significant variability associated with the rotation which suggests that large portion of the stellar surface is covered by activity complexes — spots. The detailed analysis of the connection between superflares and spots is performed in Maehara et al. (2017). The total energy of a flare is naturally related to the duration of the event. And finally, very small (0.2--0.3\%) fraction of solar type stars manifest superflares.

\begin{figure*}[!th] 
\centering
\vspace*{-0.2cm}
\includegraphics[width=\linewidth]{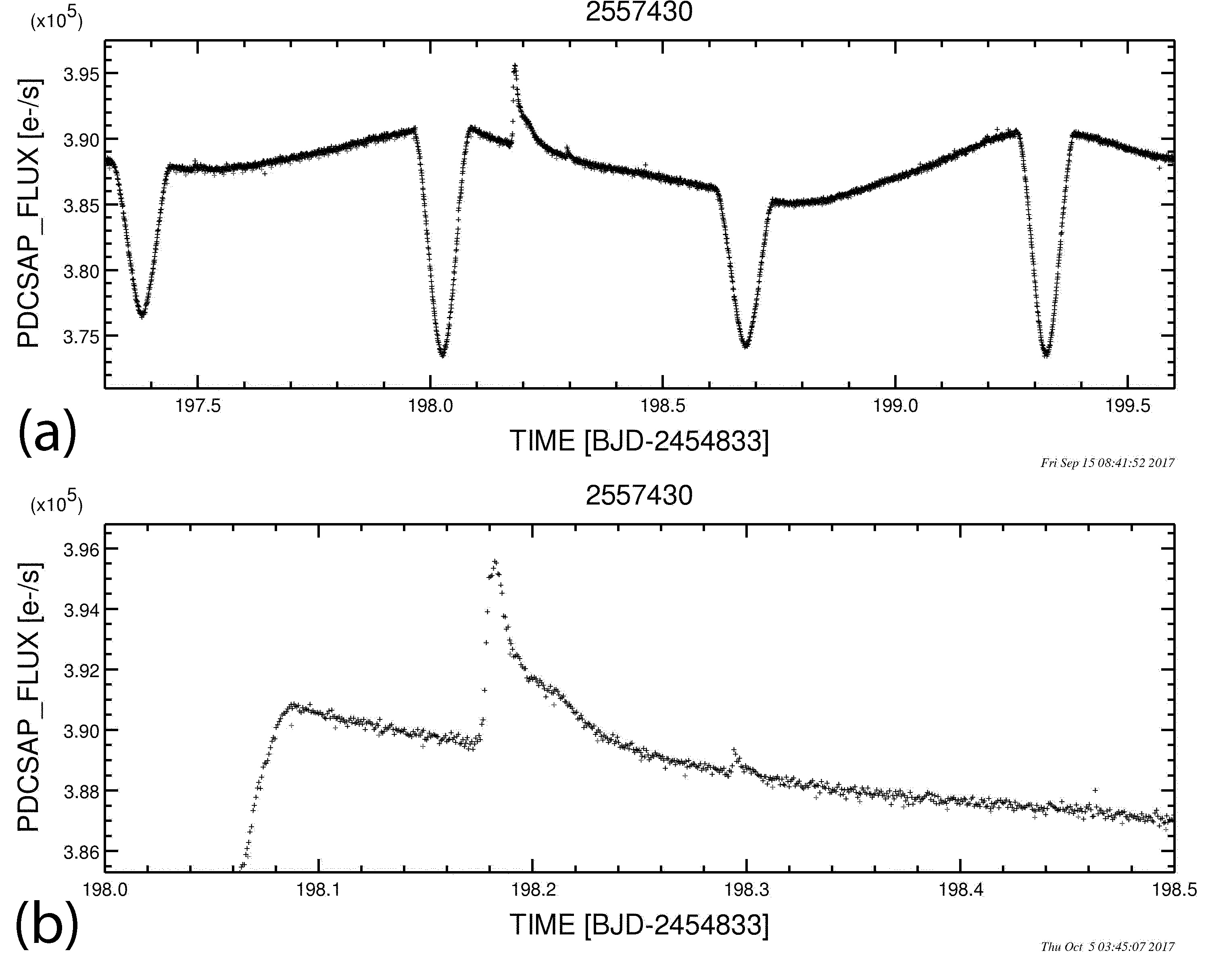}\\
\caption{
(a). The light curve of an eclipsing binary (type EA)  KIC 2557430.\\
(b). Enlarged fragment of the light curve from Fig.\,1(a) in the region of two flares. 
The energy of the stronger flare 1 near the time 198.18 is estimated as $E_{\textit{flare}} = 8.38\times 10^{35}$\,erg. 
Flare 2 near the time 198.3 has the total energy $E_{\textit{flare}} = 2.36\times 10^{34}$\,erg.
}
\label{Figure1}
\end{figure*}

We emphasize that, from the energetic considerations, the upper limit of the total flare energy on the present-day Sun is $3\times 10^{32}$\,erg, while, considering the data on the magnetic fields on young main sequence G stars with $P_{rot}$ about 10 days and the age of order 1 Gyr, the maximum flare energy there cannot exceed $10^{34}$\,erg (Livshits et al., 2015, Katsova and Livshits, 2015). Only for such phenomena one can argue that their nature is similar to that in solar flares, namely the free energy of the magnetic field is accumulated in the chromosphere and is subsequently released in the course of a nonstationary process. Therefore, in order to explain more energetic events, one should consider other mechanisms, starting with a more thorough analysis of the stars where such superflares were detected.

\section{Stars where superflares with the total energy $E > 10^{35}$\,erg were registered}

The data with the time resolution of 1 min suit the best for investigation of flare time profiles which is necessary for understanding their physical nature. These data are at hand only for 4828 stars for several months. The first evidence for a strong correlation between the flare energy, the star luminosity and radius is given in Balona (2015). It is shown that long-duration flares occur in stars with low gravity. Moreover, for three eclipsing binaries with large number of flares no relation between the flare occurrence and the orbital phase is revealed which contradicts to the idea that flares in close binaries are the result of the interaction of the components’ magnetic fields.

\begin{figure*}[!t] 
\centering
\includegraphics[width=\linewidth]{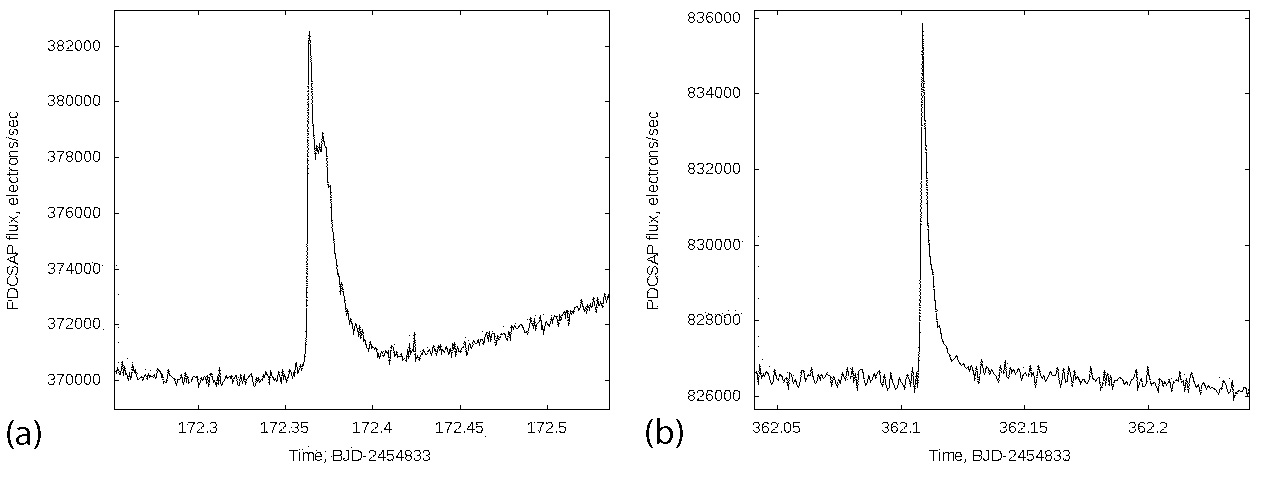}\\
\caption{The light curves of superflares in two single stars KIC 7339343 (G1) (a) 
and KIC 12072958 (F8) (b). The flare in KIC 7339343 is presented against longer-term 
variability associated with the rotation modulation of the star. The total energy 
of the flare in KIC 7339343 is $\log E_{\textit{flare}} = 35.02$ and that in KIC 12072958 
is $\log E_{\textit{flare}} = 35.229$.}
\label{rotcurve}
\end{figure*}

In article by Balona (2015), 209 stars are selected, and it is detected 3140 flares. 
Among these objects, there appear not only single stars with the variability 
caused by axial rotation, but also components of binary systems as well as 
hotter stars. We present here examples of some superflares in the light curves 
of one of eclipsing binaries, KIC 2557430, in Fig.\,1a, b, according to Kepler archive. 
The energy of the stronger flare 1 near the time 198.18 is estimated as 
$E_{\textit{flare}} = 8.38\times 10^{35}$\,erg. Flare 2 near the time 198.3 has the total energy 
$E_{\textit{flare}} = 2.36\times 10^{34}$\,erg. Among single stars we selected two stars KIC 7339343 (G1) 
and KIC 12072958 (F8) where powerful flares were detected. The light curves of superflares 
on these stars are presented in Fig.\,2. The total energy of these large flares and their duration are 
$\log E_{\textit{flare}} = 35.02$ with duration 1.128 hour and 
$\log E_{\textit{flare}} = 35.229$ with duration 0.556 hour correspondingly.

In order to understand in which stars the most powerful events occur we considered 
the objects where, according to Balona (2015), flares with the total energy from 
$10^{32}$\,erg to $3.7\times 10^{37}$\,erg were registered in. Fig.\,3 presents 
the fundamental parameters of all these stars -- effective temperatures $T_{\textit{eff}}$ 
and surface gravitational accelerations $\log g$. There are three generations of 
Kepler Input Catalog (KIC): the original KIC (Brown et al. 2011), the first update 
(Huber et al. 2014) and the second update (Mathur et al., 2017).  Balona (2015) 
used the original version of 2011; in contrast, we adopted updated atmospheric 
parameters from  Mathur et al. (2017). Mathur et al. (2017) improved the parameters 
used earlier by taking better input values and following a refined methodology 
of parameter inference from the isochrones. In fact, this Fig.\,3 is the 
Herzsprung-Russell diagram for stars considered.  

From the list of these objects we selected those which demonstrated flares 
with the total energy $E > 10^{35}$\,erg including binaries (Tables 1 and 2). 
This new list consists of 46 single F5, G and K stars which are believed to 
be of solar type ($T_{\textit{eff}} < 6500$\,K) and 22 components of detached 
(EA) and semi-detached (EB) eclipsing binary systems.  

Let us analyze in more detail the fundamental parameters of the stars where 
the most powerful flares with $E > 10^{35}$\,erg were discovered. The properties 
of the selected objects are given in Tables 1 and 2 for single and eclipsing binary 
stars respectively. It should be noted that some objects belong to other types than 
it was indicated in Balona (2015), as it follows from SIMBAD database. For instance, 
this regards to KIC 12156549 where instead of a single star it turns out to be a 
binary DA+dMe -- an oscillating white dwarf.

\begin{figure*}[!t] 
\vspace*{-1cm}
\begin{center} 
\includegraphics[width=\linewidth]{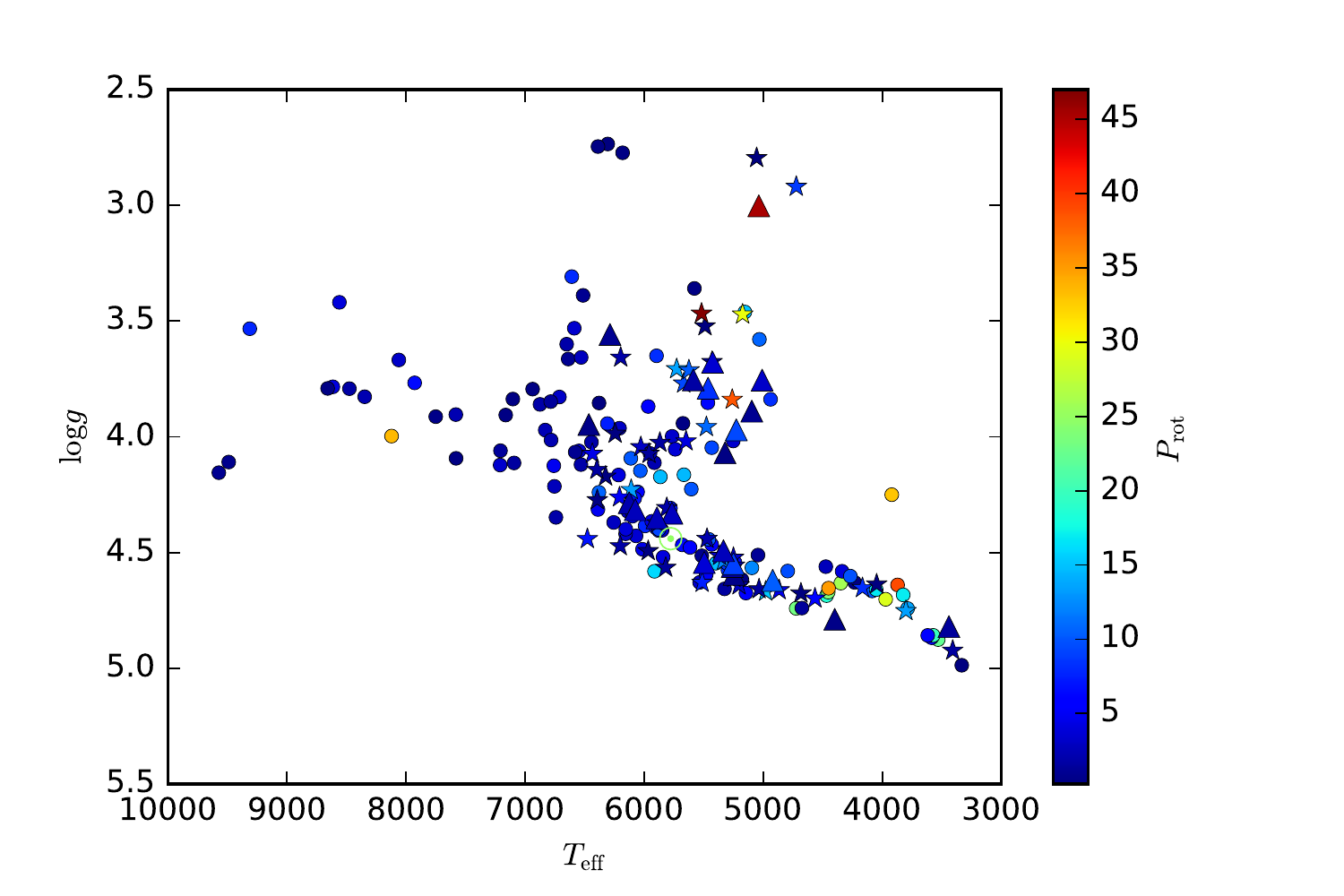}\\ 
\caption{The updated fundamental parameters -- effective temperatures $T_{\textit{eff}}$ and 
surface gravitational accelerations $\log g$ -– adopted from Mathur et al. (2017) 
for atmospheres of all 209 stars by Balona (2015). Single stars with the total flare energy 
$E_{\textit{flare}} > 10^{35}$\,erg are marked with asterisks, detached (EA) and semi-detached (EB) binaries with 
$E_{\textit{flare}} > 10^{35}$\,erg are denoted with triangles. The rest 141 stars with 
less powerful flares ($E_{\textit{flare}} \le 10^{35}$\,erg) are shown with circles. Location 
of the Sun is given by its sign. The color bar in the right encodes the rotation period of each star.}
\label{rotcurve}
\end{center}
\end{figure*}

The stars with the largest flares ($E > 10^{35}$\,erg) comprise about 30\%\ of the set considered by Balona (2015). The locations of these stars relative to the rest 70\%\ are given in the $\log g – T_{\textit{eff}}$ diagram (Fig.\,4). In this figure we also show the dependencies between $T_{\textit{eff}}$ and $\log g$ for the main-sequence stars and subgiants (luminosity classes V and IV respectively) derived from the evolutionary tracks on the basis of Straižys and Kuriliene (1981).

Fig.\,4 demonstrates that stars with the most powerful flares reside as a rule in the region between two curves representing the main sequence and the subgiant branch. Larger scatter is characteristic to the stars which are so far referred to as of “solar type”. However, from this figure it is clear that their similarity with the Sun is somewhat questionable because most of them have lower gravitational acceleration (and hence larger radii) and different effective temperatures. Some of these single stars are subgiants and even giants. In such stars, larger scales of active regions and coronal loops are favourable for occurence of longer flares and hence larger total flare energy. It is also noted that these stars do not have “hot Jupiters” as a rule, except two eclipsing binaries KIC 2162635 and KIC 4055765 which are suspected in possessing of planet (accordingly to SIMBAD database). 

To somewhat lesser extent the difference from the main-sequence parameters are 
characteristic to the stars in eclipsing binaries which have different degrees 
of axial and orbital (from less than a day to several tens of days) period 
synchronization. Among these stars there are binaries of W UMa-, RS CVn-, 
$\beta$ Lyr-, Algol-types. Balona (2015) checked 
whether flare occurrence is related to the orbital phase
on the example of three eclipsing systems with components of different types. 
It is shown that the flaring activity does not depend on the orbital phase. 
It is evidence that flares occur in the star and not in the space between 
the components.

\section{Conclusion}

Thus, most part of late-type stars with registered powerful superflares with $E > 10^{35}$\,erg turn out to be either young fast rotators or have radii larger than that of the Sun, or are members of close binary systems. The nature of such phenomena should differ significantly from the solar one since the magnetic fields observed in solar-type G-stars do not exceed a few Gauss and can only provide flares with the total energy of up to $2\times 10^{34}$\,erg. In order to account for more powerful phenomena one needs to engage some another energy reservoir as a source of the primary energy release compared to a solar flare or another regime of the dynamo mechanism. The last possibility can be realized in young stars rotating very fast and component of close binaries as well, which can possess another hydrodynamics of the inner layers inside the stars. On this way, one of possible solution of this problem is considered by Katsova et al. (2018). They find that anti-solar differential rotation or anti-solar sign of the mirror-asymmetry of stellar convection can provide strong magnetic field in dynamo models.

\begin{figure*}[!tb] 
\vspace*{-1cm}
\begin{center} 
\includegraphics[width=\linewidth]{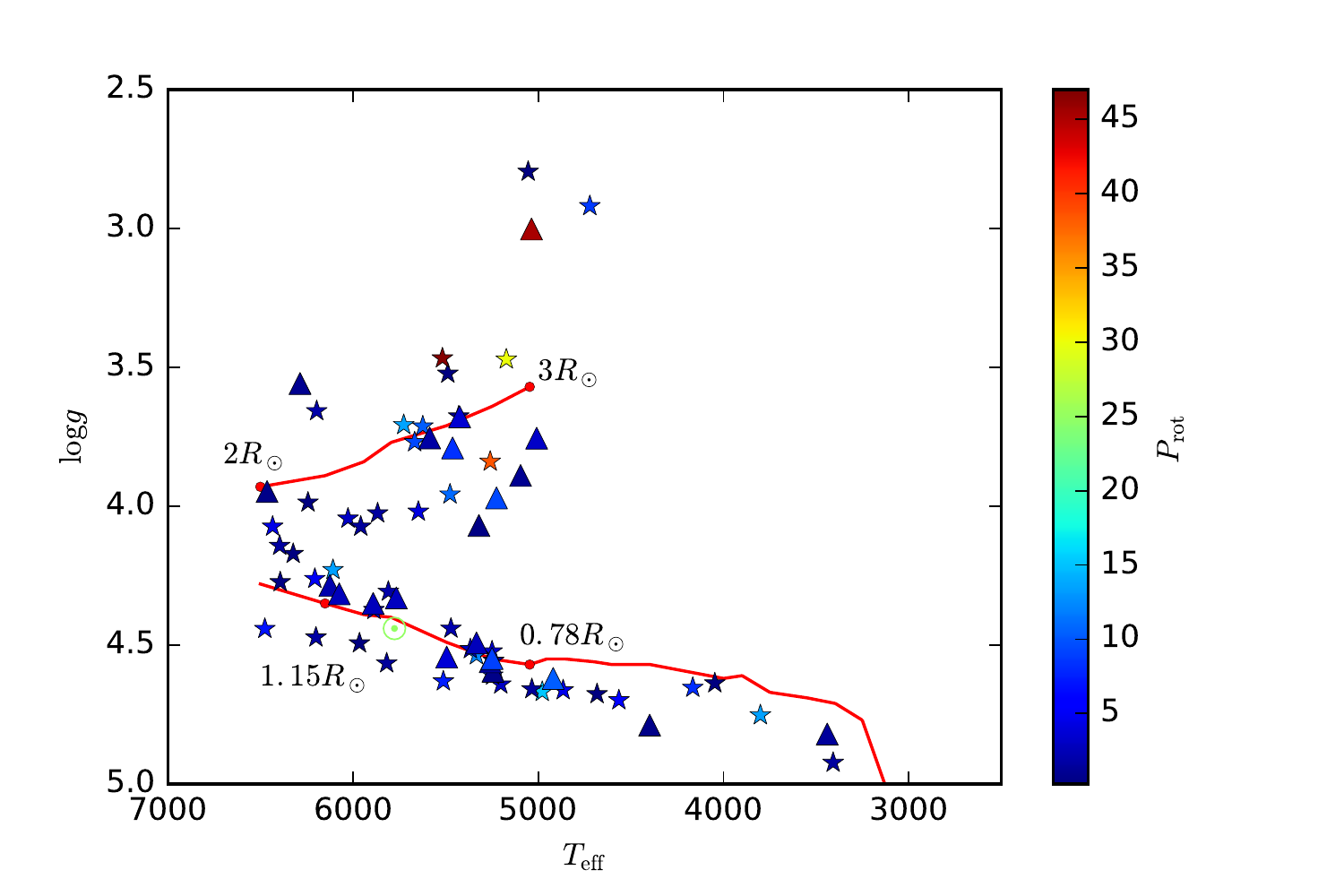}\\ 
\caption{“$\log g – T_{\textit{eff}}$” diagram with the same designations for single “solar-type” stars with $T_{\textit{eff}} < 6500$\,K, detached (EA) and semi-detached (EB) binaries with superflare total energy $> 10^{35}$\,erg . The upper solid curve corresponds to the parameters of the main-sequence stars, the lower curve -- to that of subgiants accordingly to Straižys and Kuriliene (1981).}
\label{rotcurve}
\end{center}
\end{figure*}

\phantomsection
\bibliographystyle{unsrt}

\begin{table*}

\vspace{-0.5cm} 
\caption{Parameters of single late-type stars with the maximum flare energy $\ge 10^{35}$\,erg.\\
 In the columns are given: Kepler Input Catalog ID, effective temperature, $T_{\textit{eff}}$  in K, 
 logarithm of gravitational acceleration, $\log g$, radius in $R_\odot$, rotation period, 
 and logarithm of the energy of the most powerful flare in erg.}
\vspace{0.3cm} 

\large\centering
\begin{tabular}{|c|c|c|c|c|c|} 
\hline
 KIC  &     $T_{\textit{eff}}$, K    &    $\log g$  &   Radius, $R_\odot$  &  $P_{rot}$, day  &  $\log E$ \strut\cr
\hline

2300039 & 	3408  &  4.923  & 0.325 & 1.707  & 35.18 \cr
2852961 & 	4722  &  2.919  & 5.499 & 8.8    & 38.18 \cr
3128488 & 	4565  &  4.698	& 0.546	& 6.16   & 35.31 \cr
3441906 & 	5242  &  4.556	& 0.722	& 1.853  & 35.93 \cr
3945784 & 	4979  &  4.668	& 0.588	& 15.267 & 35.09 \cr
4273689 & 	5173 & 	 3.471 & 	2.742 & 	29.943 & 	36.38 \cr
4543412 & 	5472 & 	 4.44  & 	0.877 & 	2.165  & 	35.47 \cr
4671547 & 	4166 & 	 4.653 & 	0.652 & 	8.138  & 	35.15 \cr
5475645 & 	5513 & 	 4.63  & 	0.704 & 	7.452  & 	35.63 \cr
5609753 & 	5203 & 	 4.641 & 	0.62  & 	3.162  & 	35.47 \cr
5733906 & 	5430 & 	 3.676 & 	2.365 & 	0.719    & 36.11 \cr
6437385 & 	5727 & 	 3.707 & 	2.061 & 	13.672	 & 36.78 \cr
6545986 & 	6394 & 	 4.273 & 	1.291 & 	0.557	 & 35.31 \cr
6786176 & 	6477 & 	 4.441 & 	1.035 & 	6.85	 & 35.02 \cr
7206837 & 	6324 & 	 4.171 & 	1.54  & 	-		 & 35.11 \cr
7339343 & 	5810 & 	 4.307 & 	1.136 & 	2.064 & 	35.14 \cr
7350496 & 	5668 & 	 3.769 & 	2.149 & 	9.403 & 	36.43 \cr
7420545 & 	5260 & 	 3.839 & 	1.943 & 	38.55 & 	35.79 \cr
7849619 & 	3801 & 	 4.752 & 	0.492 & 	13.55 & 	35.47 \cr
7940546 & 	6244 & 	 3.986 & 	1.807 & 	-	  & 	35.18 \cr

7944142 & 	5055 & 	 2.795 & 	11.123 & 	-	  & 	35.6 \cr
8226464 & 	6028 & 	 4.044 & 	1.535  & 	3.101 & 	36.44 \cr
8481574 & 	5966 & 	 4.493 & 	0.961  & 	0.328 & 	35.06 \cr
8651471 & 	5250 & 	 4.522 & 	0.771  & 	3.424 & 	35.45 \cr
8656342 & 	5959 & 	 4.073 & 	1.587  & 	1.434 & 	35.45 \cr
8682921 & 	5489 & 	 3.522 & 	3.51  & 	0.254  & 	35.52 \cr
8915957 & 	5518 & 	 3.467 & 	2.652 & 	46.67  & 	35.93 \cr
8957218 & 	5477 & 	 3.957 & 	1.607 & 	10.881 & 	35.81 \cr
9349698 & 	5035 & 	 4.658 & 	0.6   &     1.359  & 	35.36 \cr
9450669 & 	4866 & 	 4.662 & 	0.593 & 	4.774  & 	36.36 \cr
9652680  &  5819 & 	 4.565 & 	0.823  & 	1.408  & 	35.38 \cr
9752973  &  6109 & 	 4.229 & 	1.173  & 	13.5   &    35.03 \cr
9833666  &  5624 & 	 3.712 & 	2.702  & 	10.341 & 	36.27 \cr
10063343 & 	4047 & 	 4.637 & 	0.631  & 	0.333  & 	35.48 \cr
10068383 & 	5247 & 	 4.609 & 	0.744  & 	8.602  & 	35.64 \cr
10355856 & 	6435 & 	 4.073 & 	1.672  & 	4.487  & 	35.56 \cr
10528093 & 	5334 & 	 4.536 & 	0.746  & 	12.176 & 	36.19 \cr
10976930 & 	6197 & 	 3.657 & 	3.01   &    2.054  & 	36.32 \cr
11137395 & 	6397 & 	 4.142 & 	1.472  & 	1.57   &    35.41 \cr
11445774 & 	6201 & 	 4.472 & 	0.989  & 	1.744  & 	35.33 \cr

11551430 & 	5648 & 	 4.019 & 	1.605 & 	4.145 & 	36.87 \cr
11560431 & 	5367 & 	 4.514 & 	0.828 & 	3.142 & 	35.06 \cr
11610797 & 	5868 & 	 4.025 & 	1.67  & 	1.625 & 	35.76 \cr
11665620 & 	4683 & 	 4.676 & 	0.573 & 	0.363 & 	36.15 \cr
12072958 & 	6207 & 	 4.261 & 	1.097 & 	5.107 & 	35.23 \cr

12156549 &	5888 &	 4.373 &	1.043 &     3.651 &     36.5 \cr
 
 \hline 
\end{tabular}
\end{table*}

\begin{table*}[!t]

\caption{
 Parameters of eclipsing binary stars with the maximum flare energy $\ge 10^{35}$\,erg.\\
 In the columns are the same as in Table\,1: Kepler Input Catalog ID, effective temperature, $T_{\textit{eff}}$  in K, 
 logarithm of gravitational acceleration, $\log g$, radius in $R_\odot$, rotation period, 
 and logarithm of the energy of the most powerful flare in erg.}
 
\vspace{0.3cm} 
\large\centering
\begin{tabular}{|c|c|c|c|c|c|} 
\hline
 KIC  &     $T_{\textit{eff}}$, K    &    $\log g$  &   Radius, $R_\odot$  &  $P_{rot}$, day  &  $\log E$ \strut\cr
\hline
&&&&&\cr
2162635 &   5009 &    3.755 & 2.283 & 3.3   & 35.54 \cr
2438502 & 	5463 & 	  3.79  & 2.445 & 8.299 & 35.32 \cr
5952403 & 	5037 & 	  3.002 & 7.553 & 45.28 & 37.57 \cr
6205460 & 	5425 & 	  3.678 & 2.592 & 3.717 & 36.91 \cr
6548447 & 	5226 & 	  3.97  & 1.643 & 9.409 & 36.38 \cr
&&&&&\cr
7885570 & 	5587 & 	  3.754 & 	2.553 & 	1.73   & 	35.49 \cr
7940533 & 	5495 & 	  4.543 & 	0.798 & 	3.826  & 	36.15 \cr
8081482 & 	5767 & 	  4.33	&   1.074 & 	2.819  & 	36.45 \cr
8435247 & 	5245 & 	  4.594 & 	0.674 & 	0.696  & 	36.08 \cr
8590527 & 	6465 & 	  3.947 & 	1.847 & 	0.737  & 	35.52 \cr
&&&&&\cr
8608490 & 	5096 & 	  3.889 & 	1.782 & 	1.083 & 	36.45 \cr
8669092 & 	6287 & 	  3.558 & 	3.177 & 	0.998 & 	35.93 \cr
9328852 & 	4399 & 	  4.788 & 	0.463 & 	0.646 & 	35.45 \cr
9569866 & 	5260 & 	  4.56	&   0.717 & 	1.468 & 	35.53 \cr
9576197 & 	5250 & 	  4.548 & 	0.736 & 	9.096 & 	36.18 \cr
&&&&&\cr
9641031	 & 6126	  &  4.285 & 	1.176 & 	2.156  & 	35.88 \cr
9655129	 & 5334	  &  4.492 & 	0.81  & 	2.75   &    35.83 \cr
9705459	 & 5892	  &  4.35  & 	1.198 & 	2.796  & 	35.91 \cr
10547685 & 	6075  &  4.315 & 	1.13  & 	2.718  & 	35.3 \cr
11548140 & 	3440  &  4.82  & 	0.439 & 	1.3778 & 	35.74 \cr
&&&&&\cr
11551692 & 4920 & 	  4.62  & 	0.72  & 	10.418 & 	36.06 \cr
11560447 & 5321 & 	  4.069 & 	1.475 & 	0.526  & 	35.8 \cr

 \hline 
\end{tabular}
\end{table*}

\end{document}